\documentclass[aps,twocolumn,epsfig,graphics,superscriptaddress,showpacs,floatfix,mathbbm,nopacs]{revtex4}

\usepackage{amsmath,amsfonts,amssymb,amsthm,stmaryrd,graphics,graphicx,epsfig,color,times,bbm,psfrag}

\newcommand{\cc}{\mathbb{C}}

\newcommand{\id}{\mathbb{I}}

\begin{document}
\title{On the experimental feasibility of
continuous-variable optical entanglement distillation}

\author{J.~Eisert}
\affiliation{QOLS, Blackett Laboratory,
Imperial College London, Prince Consort Road, London SW7 2BW, UK}
\affiliation{Institute for Mathematical Sciences, Imperial College
London, Prince's Gardens, London SW7 2PE, UK}

\author{M.B.\ Plenio}
\affiliation{QOLS, Blackett Laboratory,
Imperial College London, Prince Consort Road, London SW7 2BW, UK}
\affiliation{Institute for Mathematical Sciences, Imperial College
London, Prince's Gardens, London SW7 2PE, UK}

\author{D.E.\ Browne}
\affiliation{Department of Materials, Oxford
University, Parks Road,
Oxford OX1 3PH, UK}

\author{S.\ Scheel}
\affiliation{QOLS, Blackett Laboratory,
Imperial College London, Prince Consort Road, London SW7 2BW, UK}

\author{A.\ Feito}
\affiliation{QOLS, Blackett Laboratory,
Imperial College London, Prince Consort Road, London SW7 2BW, UK}
\affiliation{Institute for Mathematical Sciences, Imperial College
London, Prince's Gardens, London SW7 2PE, UK}


\begin{abstract}
Entanglement distillation aims at preparing highly entangled
states out of a supply of weakly entangled pairs, using local
devices and classical communication only. In this note we
discuss the experimentally feasible schemes for optical
continuous-variable entanglement distillation that have been
presented in [D.E.\ Browne, J.\ Eisert, S.\ Scheel, and M.B.\
Plenio, Phys. Rev. A {\bf 67}, 062320 (2003)] and [J.\ Eisert,
D.E.\ Browne, S.\ Scheel, and M.B.\ Plenio, Annals of Physics (NY)
{\bf 311}, 431 (2004)]. We emphasize their versatility in
particular with regards to the detection process and discuss the
merits of the two proposed detection schemes, namely
photo-detection and homodyne detection, in the light of
experimental realizations of this idea becoming more and more
feasible.
\end{abstract}

\date{\today}
\maketitle

 The ability to distribute entanglement over large
distances is one of the key pre-requisites for many
practical implementations of quantum communication schemes. Quite
spectacular experimental progress has indeed been made in recent
years towards reaching this aim.  Several functioning
medium-distance quantum key distribution schemes have been
reported and successful tests indicating a violation of Bell's
inequalities have been carried out.

Needless to say, any mechanism leading to losses and
decoherence will eventually deteriorate entangled states into
merely classically correlated quantum states. Such states may
then, for example, no longer be useful in the sense that the
generation of a secure classical key cannot be
guaranteed. To regain the ability to distribute
entanglement in the presence of noise, some instance of an
entanglement distillation scheme or quantum repeaters is
required. In such entanglement distillation schemes \cite{Dist},
highly entangled states are extracted from a situation where
entanglement is present in only a dilute form. In
practical optical schemes such methods form one of the
building blocks towards making long-distance quantum communication
possible.

A number of distillation schemes have been devised for discrete
degrees of freedom of light (in particular polarization degrees of
freedom), and in some instances even been experimentally realized
\cite{Gisin,Zeil,Pan}. Considering the photon-number or
continuous-variable degree of freedom, in turn, offers an
interesting alternative to the former setting, allowing also in
principle the realization of event-ready entanglement distillation
without the need of destructive post-selection or photon counters.
In this context Gaussian states and operations are of particular
interest as they are experimentally relatively easily accessible.
It came as a surprise, yet, that with Gaussian operations alone,
continuous-variable entanglement can not be distilled
\cite{NoGo1,NoGo2,NoGo3}. This refers to Gaussian input states,
and manipulation with passive and active optical elements,
homodyne detection, and vacuum projections. Fortunately, it was
then demonstrated in Refs.\ \cite{Old,New} that one only needs to
leave the Gaussian setting in a single step to break this no-go
theorem. From there on entanglement distillation is indeed
possible, making use of Gaussian operations, in fact using passive
linear optics and vacuum projections or homodyne measurements
only.

As the realizations of the experimentally feasible schemes of
Ref.\ \cite{Old,New} is coming closer in view of the techniques
that have been developed in recent years, it seems worth
discussing and emphasizing the versatility of this approach
without presenting mathematical details. This is the aim of
this brief note. As has already been discussed in Ref.\
\cite{Old}, photon detectors (with relatively small detection
efficiency) are suitable in the iteration of the scheme, as well
as homodyning techniques. Under no circumstances, highly efficient
photon counters with photon number resolution are required.

\begin{figure}[ht]
\center{\includegraphics[scale=0.65]{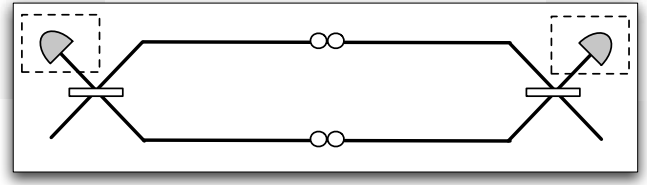}}
 \caption{One step of the procedure. The dotted
 rectangle represents the local detection.}\label{Scheme}
 \end{figure}

\begin{figure}[ht]
\center{\includegraphics[scale=0.5]{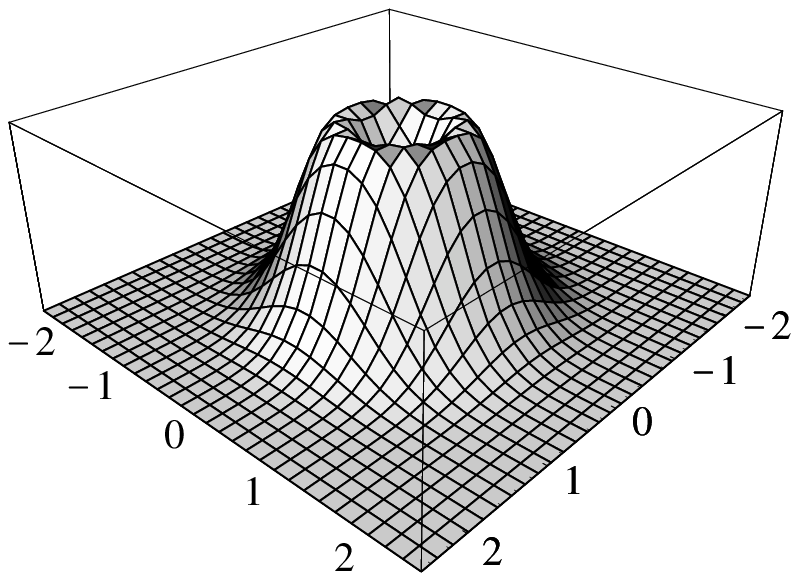}}

\center{\includegraphics[scale=0.5]{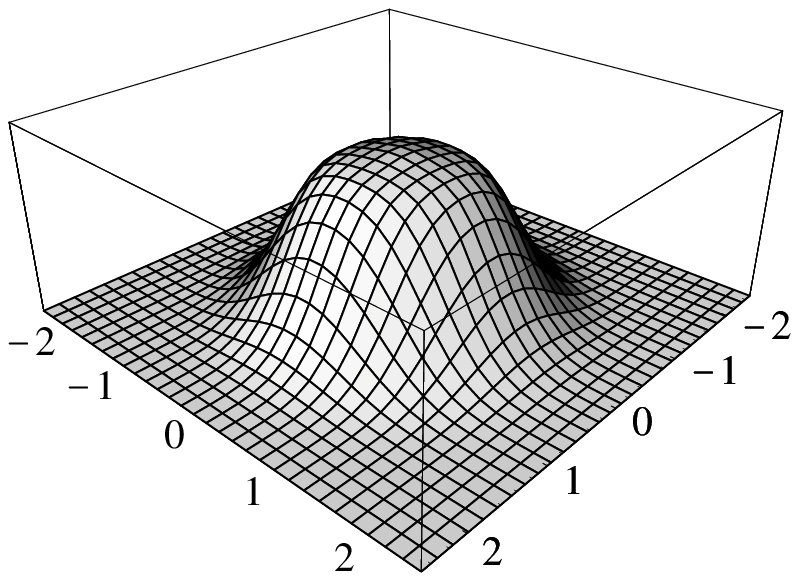}}

\center{\includegraphics[scale=0.5]{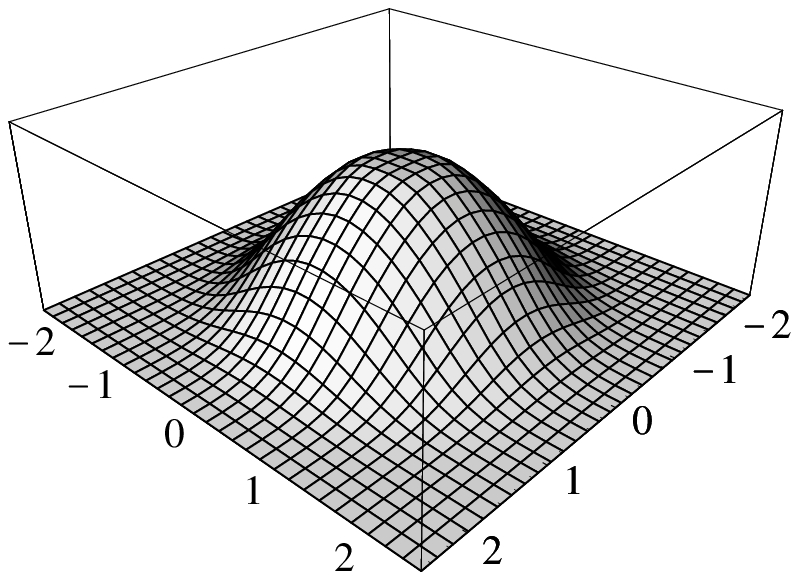}}

 \caption{An instructive plot of the scheme used as a
 single-mode procedure:
 this plot depicts the Wigner function of the single-mode
 state after zero, one, and two steps of the procedure.
 The initial state is taken to be $|0\rangle + \varepsilon |1\rangle)/
 (1+\varepsilon^2)$, where again $\varepsilon=0.95$
 (taken from Ref.\ \cite{New}).}\label{Wigner}
 \end{figure}

{\it The iteration. -- } In the following we discuss the
basic iterative scheme for entanglement distillation of general
non-Gaussian states developed in Refs.\ \cite{Old,New}
(Non-Gaussian here means that the states are non-Gaussian in the
photon number degree of freedom; hence, the Wigner function of the
states is non-Gaussian \cite{Survey,Loock}). Two copies of a
non-Gaussian two-mode state $\rho$ serve as inputs to the
procedure \cite{Alvaro}. These inputs are mixed locally at a
$50/50$ beam splitter each, represented by $U_{\text{BS}}$,
leading to
\begin{equation}
    \rho'= (U_{\text{BS}}\otimes U_{\text{BS}})
    (\rho\odot \rho) (U_{\text{BS}}\otimes U_{\text{BS}})^\dagger.
\end{equation}
Here, the symbol $\odot$ denotes a tensor product with respect to
the two identical two-mode states. In turn, $\otimes$ is meant as
the tensor product between the two remote parties (see Fig.\
\ref{Scheme}). Then, one of the outputs of each of the beam splitters
undergoes a Gaussian measurement. The outputs of the other arms of
the beam splitters are retained, in case of 
a successful measurement
event, and are then used as the input of the next step of the
procedure. This can be done in an iteration, and each instance
will from now on be referred to as {\it one step of the procedure},
or a ``Gaussification step''. One such step is presented in Fig.\
\ref{Scheme}. In practice, already a single step alone can
significantly increase the degrees of entanglement, and this is the
setting that seems indeed realistic to reach with present
technology. In Refs.\ \cite{Old,New}, the action of such an
iteration on {\it all} non-Gaussian two-mode states has been
studied in detail. In particular, (i) a proof of convergence to
Gaussian states has been delivered for all pure or mixed initial
states (compare also Fig.\ \ref{Wigner}), (ii) the increase of the
degree of entanglement has been studied, (iii) the increase in the
degree of squeezing, and (iv) necessary and sufficient criteria
have been presented to decide when an exact purification of the input state is possible. 
In the single mode case, enhancement of squeezing
and the loss of non-classical features has been discussed. The
action of the scheme under one or a few steps has also been
studied in detail.

The non-Gaussian states can be thought of as resulting from a
number of possibilities; in Refs.\ \cite{Old,New}, in particular,
an idea has been explored that employs the use of photon
subtraction. Very much related steps for the creation of a
non-Gaussian state have already been demonstrated
experimentally \cite{Grangier,Heersink,Polzik,NewGrangier}.

\begin{figure}[ht]
\center{\includegraphics[scale=0.75]{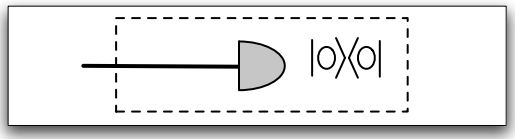}}
 \caption{Photon detection variant of one of the two
 necessary measurements.}\label{Direct}
 \end{figure}

{\it Photon detection variants. --} To realize the feasible
distillation scheme of Refs.\ \cite{Old,New} measurements have to
be carried out, retaining the outcomes corresponding to a Gaussian
projection. For this key element of the procedure a number of
approaches are possible. In one variant, the Gaussian projection
is obtained by projecting onto the vacuum, making use of photon
detectors which can discriminate between the presence and absence
of photons. Such a device --  which we assume to be perfectly
functioning for simplicity, an assumption that can easily be
relaxed  -- can be theoretically described as implementing a
measurement with Kraus operators $|0\rangle\langle0|$ (``no
click'') and $\id - |0\rangle\langle0|=\sum_{n=1}^{\infty}
|n\rangle\langle n|$ (``click''), see Fig.\ \ref{Direct}. One step
of the procedure, in a successful event of a vacuum projection,
hence amounts to the transformation of two two-mode input states
$\rho$ into
\begin{equation}
    \rho''=
    \langle 0|\otimes\langle 0 |(U_{\text{BS}}\otimes U_{\text{BS}})
    (\rho\odot \rho) (U_{\text{BS}}\otimes
    U_{\text{BS}})^\dagger |0\rangle \otimes |0\rangle/N
    \label{detectoutcome}
\end{equation}
where $N$ is an appropriate normalization.

\begin{figure}[ht]
\center{\includegraphics[scale=0.45]{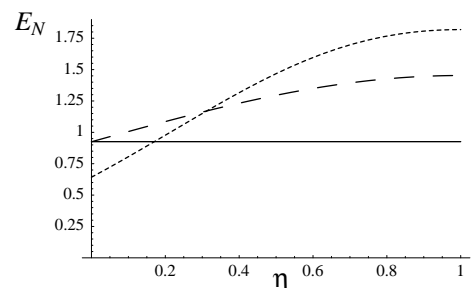}}
 \caption{ The logarithmic negativity as a function
 of the detector efficiency $\eta$ after one (dashed line) and
 after 10 steps (dotted line) of the procedure, for the initial state
 given in eq.\ (\ref{state}). The  solid line represents the logarithmic
 negativity of the initial state prior to the implementation of the
 procedure.}\label{Efficient}
 \end{figure}

This state will have a higher degree of entanglement, and may or
may not be used as the input of the next step of the procedure.
Note that this scheme does not require the ability to count
photons, but merely requires photon detectors discriminating
between presence or absence of photons. This is the variant in
terms of which most of the results in Refs.\ \cite{Old,New} have
been stated. In Ref.\ \cite{New}, the implications of detection
inefficiencies are discussed in great detail. Quite surprisingly,
the scheme is robust with respect to low detection efficiencies,
lower than the ones that are already available with present
technology. The degree of entanglement, measured in terms of the
logarithmic negativity $E_N$ as a measure of entanglement
\cite{logneg}, after a number of steps of the procedure, as a
function of the detection efficiency $\eta$ ($\eta=1$ corresponds
to perfect detectors) is depicted in Fig.\ \ref{Efficient}. The
initial two-mode state $\rho$ is taken in this plot to be
$\rho=|\psi\rangle\langle\psi|$ with
\begin{equation}
    |\psi\rangle = (|0,0\rangle + \varepsilon |1,1\rangle)/(1+\varepsilon^2),
    \label{state}
\end{equation}
for $\varepsilon=0.95$. Other examples are discussed in
Ref.\ \cite{New}.

\begin{figure}[ht]
\center{\includegraphics[scale=0.75]{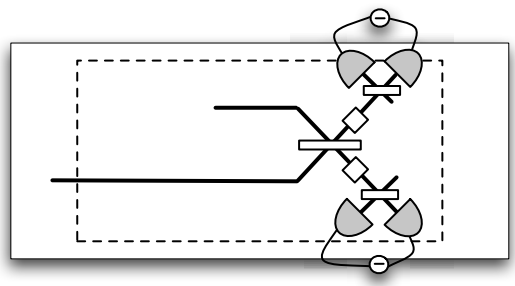}}
 \caption{Balanced homodyne detection variant
 of one of the two necessary measurements.}\label{Homodyne}
 \end{figure}

{\it Homodyne detection variants. --} As has been discussed
in Ref.\ \cite{Old}, homodyne detection is a feasible alternative
to photo-detection in the implementation of the basic protocol
detailed above. The advantage of such an approach is that the
detection efficiencies are higher for such measurement schemes
than for photon detection.

Of particular interest here is balanced homodyne detection, see
Fig.\ \ref{Homodyne}. In homodyne detection, the signal field is
combined via a beam splitter with a reference field, referred to
as local oscillator. In balanced homodyne detection with a
sufficiently strong local oscillator field, field quadratures can
be measured. In the scheme depicted in Fig.\ \ref{Homodyne},
having an additional  input, followed by passive linear
optics, and homodyne detection, the above measurement can indeed
be realized. This setup is in fact the familiar setup to measure
to $Q$-function of a single mode. Concerning the choice of the
passive optics, see also the appendix.
This has been discussed in
the context of entanglement distillation in Ref.\ \cite{Old}, but
is as such a well-known observation on general measurement schemes \cite{Ulf,Vogel,Schleich}.
Refs.\ \cite{NoGo1,NoGo2,NoGo3}
discuss the possibility of using homodyne detection in
order to realize projections onto Gaussian states in a
deterministic manner in the language of covariance matrices,
i.e., moments of quadrature operators. Note that other 
amplification schemes, even phase insensitive
amplification, would in principle also be suitable.

Balanced homodyne detection leads to the implementation of a
projection on a coherent states $|\alpha\rangle$ or, in other
words, a measurement with POVM elements
\begin{equation}
    \{ |\alpha\rangle\langle \alpha|/\pi : \alpha\in \cc\}.
\end{equation}
As a consequence, such a 
balanced homodyne detection leads us to replace
the expression eq.\ (\ref{detectoutcome}) by
\begin{equation}
    \rho''=
     \langle\alpha|\otimes\langle\beta |(U_{\text{BS}}\otimes U_{\text{BS}})
    (\rho\odot \rho) (U_{\text{BS}}\otimes U_{\text{BS}})^\dagger
    | \alpha\rangle\otimes|\beta\rangle/N
\end{equation}
 for any complex $\alpha,\beta$ and, again, for appropriate
normalization $N$. In the language of moments, each of these
projections refers to one onto a single-mode Gaussian state with
second moments given by $\gamma=\id$, and first moments
$d=(\text{Re}(\alpha), \text{Im}(\alpha))$. If one now accepts
only those measurements that correspond to complex $\alpha$ and
$\beta$ close to the origin -- effectively introducing a cut-off--
it is effectively as if the original vacuum projection
has been implemented. That is, one would for
a given $x>0$ only accept outcomes for which in phase
space $|\alpha|<x$. For small $\alpha$, we can
approximate $|\alpha\rangle = |0\rangle + \alpha |1\rangle
+O(\alpha^2)$, so higher order contributions would be
orthogonal (in Hilbert space)
to an arbitrarily good approximation, 
depending on $x$. See also Ref.\ \cite{Freyberger} for an
effective realization of vacuum projections with homodyne
detection. So for the purposes of entanglement distillation, these
approaches are equivalent. With no exception, all results
concerning the convergence in the iteration towards Gaussian
states, the increase in the degrees of 
entanglement and squeezing
are just as applicable as in the previous variant, without
modification \cite{Old,New}, as the postselected state is
identical to arbitrary approximation under appropriate filtering.

There is a trade off between the achieved rate when filtering
successful outcomes and the quality of the distilled state: a too
weak filtering has the same effect as having non-unit detection
efficiencies in the previous variant, see above. The
interesting feature of this variant is -- as has been pointed out
in Ref.\ \cite{Old} -- that the high detection efficiencies of
homodyning techniques can be made use of. 
The disadvantage may be
a lower rate in the full scheme due to filtering.

{\it Remarks on other homodyning variants. --}
Finally, it is worth mentioning that 
a direct homodyne detection of one of the output modes
of the two parties will again lead to  
the same predictions under appropriate filtering. 
The action of a homodyning 
measurement is up to displacements 
the same as a local squeezing, followed by a 
projection onto the vacuum state, in the idealized limit
of infinite squeezing. For any real non-zero
squeezing parameter $s$, we 
have that 
\begin{equation}
	[U_{\text{BS}}, S(s)\odot S(s)]=0,
\end{equation}
and hence,
\begin{eqnarray}
		\langle 0| (S(s)\odot \id)U_{\text{BS}}=
		\langle 0| (\id\odot S(s)^\dagger) U_{\text{BS}}
		(S(s)\odot S(s)).
\end{eqnarray}
This is, for any finite squeezing parameter $s$, 
the situation including squeezing is the same as
if the input state had been appropriately squeezed,
followed by an inverse squeezing of the final
outputs. Again, if one does a homodyne detection
and filters with respect to outcomes close to the
origin, the results stated
in Ref.\ \cite{Old,New} concerning a general
convergence to Gaussian states can be applied. Also, 
all statements in an increase
of entanglement are valid in the same manner.

{\it Non-Gaussian steps. --}  The above described procedure
will take non-Gaussian input
and yield an output that is a state that is
closer to being Gaussian. In the experimental preparation of
the non-Gaussian states, there is a large variety of possible
approaches. The scheme is sufficiently versatile to be applicable
as such to all non-Gaussian states. If one encounters a Gaussian
input from a lossy Gaussian channel, modeling photon loss and
thermal noise \cite{Channels}, then photon subtraction-type
\cite{Grangier,Sub} is one possible reasonable step
\cite{Old,New}, making use of photon detectors and using the
outcome associated with $\id-|0\rangle\langle 0| =
\sum_{n=1}^\infty |n\rangle\langle n|$  but any operation
yielding a non-Gaussian state is acceptable. A fair 
figure of merit for an assessment  of the quality of the
procedure would than  be  the increase of the degrees of
entanglement or purity of the channel output compared to the final
output of the scheme. In  special  instances, one may also
 suffer mere classical ignorance, due to classical
displacements in phase space due to phase noise. This kind of
phase noise is related to the so-called classical noise channel
\cite{Channels}. This also leads to mixed non-Gaussian states, in
case the classical weight of the mixing is non-Gaussian. Again,
the above procedure would be applicable. In cases where the
classical information about the random displacements could in
principle be retrieved, however, it may be advantageous to
directly correct for such errors compared to resorting to
entanglement distillation. In any case, the entanglement 
after Gaussification  will typically be higher than the one of
the mixed state before Gaussification (but generally  smaller
 compared to the state before mixing).

Notably, there are still many challenges to be overcome in a full
experimental realization of such an idea: mode matching at the two
beam splitters is definitely an issue, which points towards the
requirement of realizing the scheme entirely within fibers, to
avoid coupling losses \cite{Browne}. The
scheme is obviously also demanding in that a number of squeezers
with the capability to create large degrees of squeezing is
required.
Dark counts and inefficient detectors can to some extent be
harmful, the former of which may be significantly diminished by
appropriate temporal gating. All these obstacles constitute a
challenge to the experimental realization, but do not render it
prohibitively difficult. After all, all ingredients of this scheme
have already been experimentally realized.

To summarize, in this note we have emphasized the experimental
feasibility of the scheme presented in Refs.\ \cite{Old,New}. It
is notably sufficiently versatile to allow for homodyne detection
techniques as measurement schemes, instead of photon detection.
This  may be advantageous when it comes to exploiting
higher detection efficiencies, together with appropriate
filtering. This  alternative was  discussed already in
Ref.\ \cite{Old}, and holds as well for the results of the later
Ref.\ \cite{New}. Yet, with experimental implementations becoming
 increasingly  feasible, this point seems worth
emphasizing. It is the hope that this note fosters further
 experimental work in this direction.

{\it Acknowledgements. --}
This work has been supported by the DFG
(SPP 1116, SPP 1078), the EU (QAP, QUPRODIS),
the Microsoft Research Foundation, the EPSRC 
(ARF of SS, and the QIPIRC),
Merton College Oxford,
and the EURYI award of JE.

{\it Appendix. --}
In this appendix, we represent the arguments of
Refs.\ \cite{NoGo1,NoGo2,NoGo3} for realizing the above
POVM elements. In one case, we consider a bi-partite system,
consisting of parts $1$ and $2$. Part $2$ embodies exactly one mode,
part $1$ may consist of a number of $n$ modes. In this case, we consider
a projection onto the vacuum $|0\rangle\langle 0|$ in system $2$.
In the other case, we take a tri-partite system. Here, $3$ is initially prepared in $|0\rangle\langle0|$,
corresponding to an empty port.
Now $2$ and $3$,
each
consisting of one mode, undergo a certain
beam splitter transformation. Then they
are measured with homodyne detection.

It is not difficult to see that the
resulting transformation of system $1$ is identical, up to displacements
in phase space. The beam splitter transformation for this to be true is specified
by a matrix $S$ acting on the quadratures
$(x_3,p_3,x_2,p_2,x_{1,1},p_{1,1},...,x_{1,n},p_{1,n})$ of the tri-partite system.
This matrix is found to be
\begin{equation}
    S=\left[
    \begin{array}{ccccc}
    a & .& .& a & . \\
    . & a& -a & . &.  \\
      .& a& a &  .& . \\
    - a & .& .& a & . \\
   .& .& .& .  & \id_{2n}
    \end{array}
    \right],
\end{equation}
where $a=1/\sqrt{2}$. When acting on Gaussian inputs, for example, both
schemes will result in a transformation of the covariance matrix
\begin{equation}
    \gamma_{1,2} =\left[
    \begin{array}{cc}
    A_2 & C_{1,2}\\
    C^T_{1,2} & B_1
    \end{array}
    \right]
\end{equation}
of the system $1$ and $2$, according to \cite{NoGo1}
\begin{equation}
    \gamma_1'= B_1 - C^T_{1,2} (A_2+\id_2)^{-1} C_{1,2}.
\end{equation}
The covariance matrix, in turn, is defined for a state $\rho$
centered at the origin as
the matrix with entries
\begin{equation}
    (\gamma_{1,2})_{j,k}= 2 \text{Re}
    \text{tr}[\rho O_j O_k ],
\end{equation}
where $O=(x_2,p_2,x_{1,1},p_{1,1},...,x_{1,n},p_{1,n})$. For a more
detailed analysis, see Refs.\ \cite{NoGo1,NoGo2,NoGo3}.

\end{document}